\documentclass[twocolumn,showpacs,amsmath,amssymb]{revtex4}

\usepackage{graphicx}
\usepackage{dcolumn}
\usepackage{bm}
\newcommand{\etal}{\emph{et al.}}
\newcommand{\be}{\begin{equation}}
\newcommand{\ee}{\end{equation}}
\newcommand{\bfig}{\begin{figure}}
\newcommand{\efig}{\end{figure}}
\newcommand{\incl}{\includegraphics}

\begin{document}      

\title{Strongly nonlinear magnetization above $T_c$ in $\rm Bi_2Sr_2CaCu_2O_{8+\delta}$.
} 
\author{Lu Li$^1$, Yayu Wang$^1$, M. J. Naughton$^2$, S. Ono$^3$, 
Yoichi Ando$^3$, N. P. Ong$^1$
}
\affiliation{
\mbox{$^1$Department of Physics, Princeton University, New Jersey 08544, U.S.A.}\\
\mbox{$^2$Department of Physics, Boston College, Chestnut Hill, Massachusetts 02467, U.S.A.}\\
\mbox{$^3$Central Research Institute of Electric Power Industry, 
Komae, Tokyo 201-8511, Japan}
}

\date{\today}      

\begin{abstract}
Using high-resolution magnetometry we have investigated in detail the 
magnetization $M$ above the critical temperature $T_c$ in $\rm Bi_2Sr_2CaCu_2O_{8+\delta}$.
In a broad range of temperature $T$ above $T_c$, we find that $M(T,H)$ is strongly non-linear in the field $H$.
We show that as $T\rightarrow T_c$, the susceptibility $\chi(T,H)$ diverges to very large values 
($\chi \rightarrow$ -1) if measured in weak $H$.  In addition, $M(H)$ displays an 
anomalous non-analytic form $M\sim H^{1/\delta}$ in weak fields with a 
strongly $T$-dependent exponent $\delta(T)$.  These features strongly support the 
proposal that, above $T_c$, the pair condensate survives to support 
significant London rigidity.
\end{abstract}

\maketitle                   
\section{Introduction}
The notion that superconductivity in the cuprates is destroyed by thermally created vortices
has gained substantial experimental support.  In single crystals, evidence for vortex 
excitations that persist above the critical temperature $T_c$ to a temperature $T_{onset}\sim$ 
120 K has been obtained from the Nernst effect~\cite{Xu}.  An enhanced diamagnetic 
signal that scales accurately as the Nernst signal 
has also been observed using high-field magnetometry~\cite{Wang05}.  These results reveal
that, above $T_c$, significant pair-condensate 
amplitude survives up to intense magnetic fields $\bf H$.  Further, in ultra-thin films, the kinetic
inductance has been observed to persist above $T_c$~\cite{Corson}.  The sensitivity of 
superconductivity to phase fluctuations in cuprates has been investigated theoretically by
several groups~\cite{Kivelson,Franz}. 

In low-$T_c$ superconductors, the condensate vanishes above $T_c$, except inside 
evanescent droplets created by amplitude (or Gaussian) fluctuations.  The 
diamagnetic susceptibility $\chi$ from Gaussian fluctuations is very 
small ($\chi \sim 10^{-5}$)~\cite{Gollub}.  Given the vortex scenario in cuprates, 
the magnetization $\bf M$ measured above $T_c$ ought to be qualitatively different from the Gaussian picture.  
However, resolving the signal in tiny crystals is very challenging, and the
available results are confusing~\cite{Vidal,Caretta,Naughton,Lascialfari}.  
An experiment performed on aligned grains~\cite{Vidal} claims agreement with the 
mean-field Gaussian theory.  Other experiments are seemingly consistent with the 
vortex scenario, but uncover anomalies that are hard to interpret~\cite{Caretta}.  
Here we report detailed measurements in $\rm{Bi_2Sr_2CaCu_2O_8}$ (Bi 2212)
showing that the weak-field susceptibility $\chi$ diverges exponentially consistent with
Kosterlitz-Thouless (KT) behavior.  The strong nonlinearity observed in $\chi$ 
suggests the existence of a subtle rigidity in the pair wave function $\hat{\Psi}$ 
which persists above $T_c$.

\begin{figure}              
\incl[width=8.5cm]{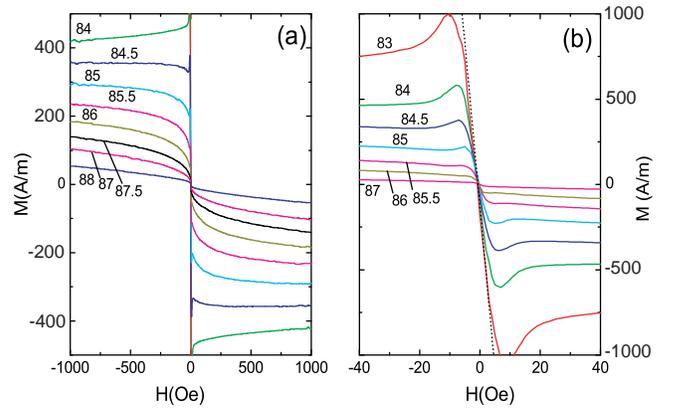}
\caption{\label{squid} 
Magnetization $M$ vs. $H$ measured in Sample 1 in fields $|H|<$ 1,000 Oe (Panel a)
and in fields $|H|<$ 40 Oe (b).  $M$ is measured in the SQUID magnetometer
in the long-time averaging mode ($\sim$12 h per curve).  In Panel (a), the sharp notch 
near $H = 0$ for $T<T_c$ (= 86 K) is $H_{c1}$.  These
features are shown more clearly in Panel (b) with a 25-fold expansion in the field
scale.  The dotted line is $M/H = \chi/(1+N_c\chi) = 2.55$ with $\chi = -0.95$ 
and $N_c$ = 0.64 for Sample I with $\bf H||\hat{c}$.
The peaks locate $H_{c1}$.
}
\end{figure}

\begin{figure}              
\incl[width=7.5cm]{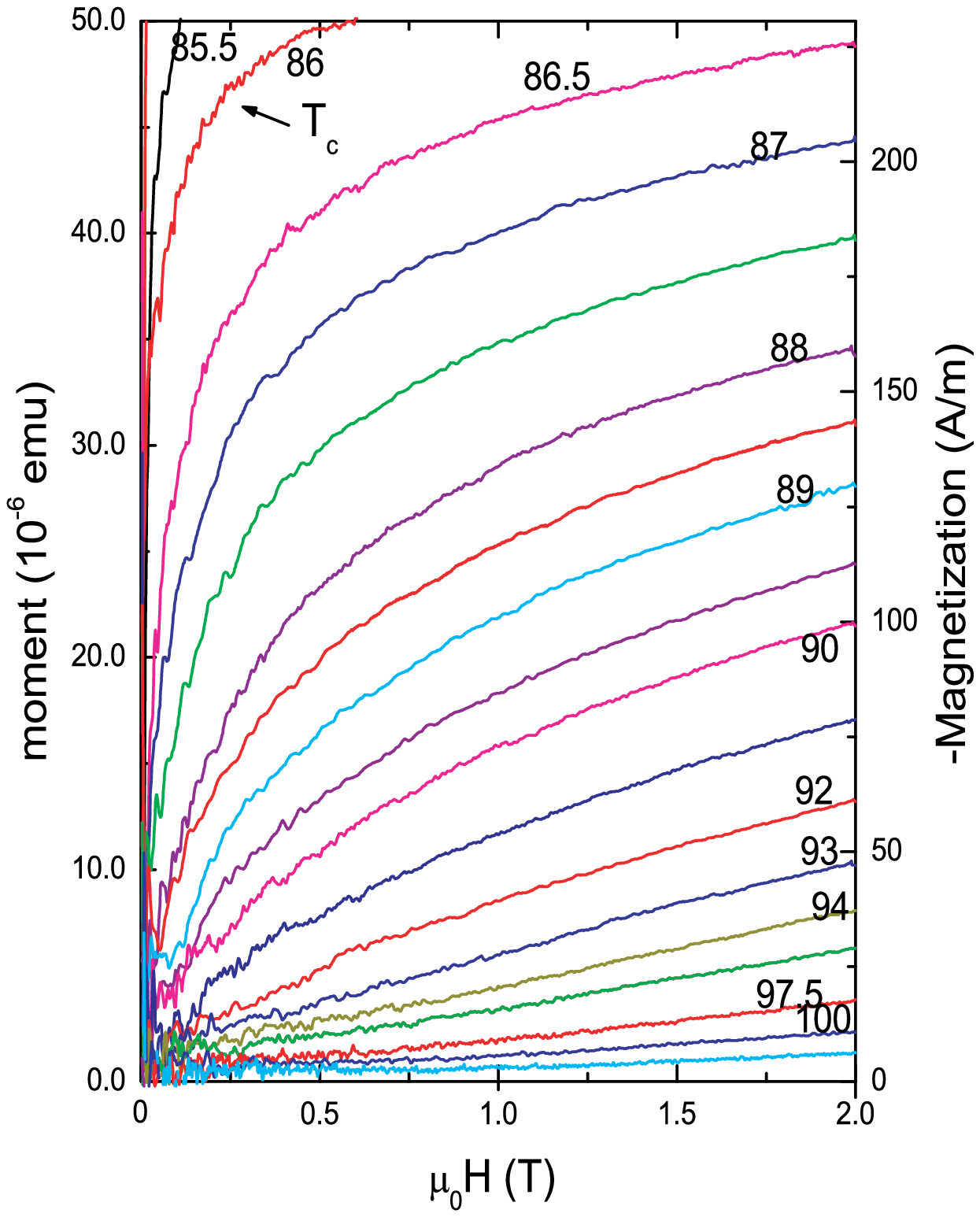}
\caption{\label{MH} 
Curves of $M$ vs $H$ in Bi 2212 (Sample I) measured by torque magnetometry
for $T$ above $T_c$ = 86 K.  The pronounced curvature evident in Fig. \ref{squid}a
persists to $\sim$100 K in fields upto 2 T and higher.  The maximum beam deflection angle
$\delta\varphi$ is 0.15$^{\rm o}$, or 10$^{-2}\varphi$.  
}
\efig

\bfig	
\incl[width=7.5cm]{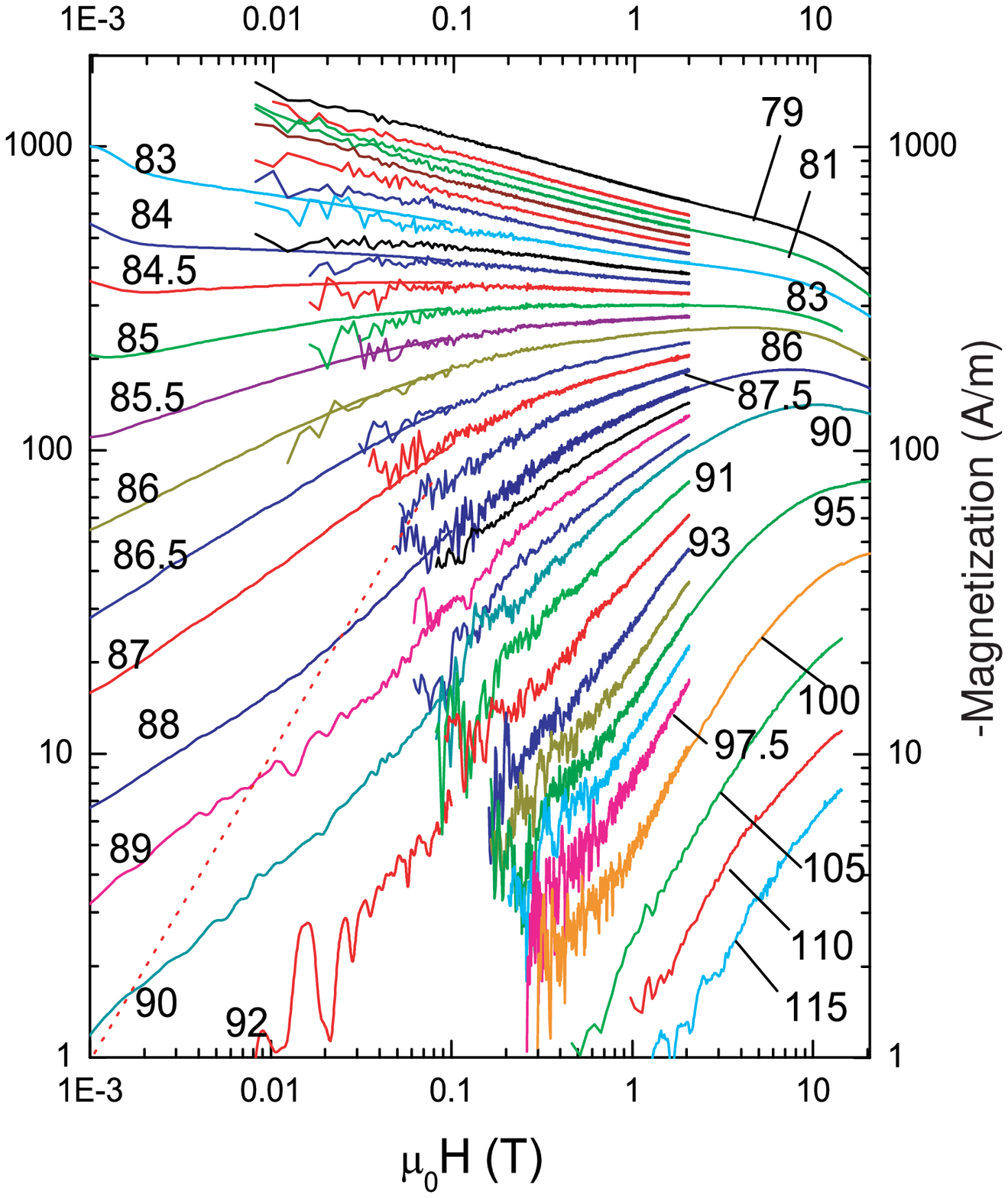}
\caption{\label{MHlog} The field dependence of $M(T,H)$ in Bi 2212 ($T_c$ = 86 K) 
from $H$ = 10 Oe to 20 T at $T$ = 79-115 K (Sample I).  The log-log plot shows that 
$M(H)$ obeys Eq. \ref{delta} as $H\rightarrow 0$.  The slight upturn 
at 83 and 84 K near 10 Oe reflects $H_{c1}$ (Fig. \ref{squid}b).
The plot combines SQUID results (10 to 1000 Oe) and toque magnetometry 
results (500 Oe to 2 T).  Torque results up to 20 T are also shown at selected $T$.  
The dashed line (of slope 1) is the linear response
of Eq. \ref{MKT} at 87 K.
}
\end{figure}

\section{Experimental details} To cover a broad range of fields, we have combined
torque magnetometry with SQUID magnetometry.
In the torque experiment, the crystal is glued at the tip of the 
Si cantilever beam with its $c$-axis at a tilt-angle $\varphi\simeq\;15^{\rm o}$ to 
$\bf H$.  In our experiment, the torque $\vec{\tau}$ is the sum of a paramagnetic term from the spin moment ${\bf m}_p$ and a 
diamagnetic term from the magnetization ${\bf M}$ of interest here, 
viz.  $\vec{\tau} = [{\bf m}_p + V{\bf M}]\times \mu_0{\bf H}$, 
with $V$ the sample volume and $\mu_0$ the permitivity.  For temperature $T>T_c$, ${\bf M}$ is 
strictly along the $z$-axis (we take $\bf \hat{z}||\hat{c}$).  An in-plane component $M_x$ is 
resolvable only below the irreversibility line~\cite{Bergemann} (from hereon, 
$M_z = M$).
We define the effective magnetization $M_{eff}\equiv\tau/(\mu_0H_xV)$ 
where $H_x = H\sin\varphi$.  
For $\varphi\ll 1$, we have 
\be
M_{eff}(T,H) = \Delta\chi_pH_z + M(T,H_z),
\label{Meff}
\ee
where the anisotropy $\Delta\chi_p = \chi_c-\chi_a$ is the difference between 
the $c$-axis and in-plane spin susceptibilities.  As explained in Ref. \cite{Wang05}, 
the very weak $T$ dependence of $\Delta\chi_p$ in cuprates allows the strongly 
$T$-dependent $M(T,H)$ to be extracted reliably from $M_{eff}$.  
All curves reported here are \emph{fully reversible} (hystereses caused by pinning of 
vortices are only seen below $\sim$50 K).  We studied 2 slightly 
underdoped Bi 2212 crystals with $T_c$ = 86 and 85 K (Samples I and II) with very similar results.

If $M$ is nearly linear in $H$ (valid if $T>1.1\;T_c$ and $H>$ 0.5 T), Eq. \ref{Meff} implies 
that the torque varies sinusoidally, viz. $\tau/V = \mu_0(\Delta\chi_p-\chi)H_xH_z \equiv A_1\sin 2\varphi$, where
$A_1(T,H)\sim H^2$ and changes sign above $T_c$, as observed in cuprates~\cite{Bergemann} (see below).

\section{Magnetization and susceptibility} 
First, we discuss the curves of $M(H)$ below $T_c$.  The curves of $M(H)$ in Fig. \ref{squid}
were measured by SQUID magnetometry.
Much analysis have focussed on the curve of $M(H)$ at 
the ``crossing temperature" $T_s$ (= 84 K) at which $M(H)$ is very nearly $H$ independent.  
By detecting the flux-exclusion in very weak fields, we have 
determined that the 3D transition at $T_c$ (86 K) actually lies 2 K above $T_s$.  
In the interval $T_s$-$T_c$, a prominent anomaly of the curves of $M(H)$
is that they curve steeply towards zero as $H\rightarrow 0$ even though full flux
exclusion obtains when $H<H_{c1}$ (the lower critical field).  This 
feature may be seen in earlier reports on $M$~\cite{Kes}, 
but has not received comment.  Closer inspection reveals that $M(H)$ displays a 
notch feature which we show in expanded scale in Fig. \ref{squid}b.   
The notch corresponds to the initial entry of vortices at $H_{c1}$.  
Below $H_{c1}$, flux exclusion gives a linear $M$-$H$ variation with a steep slope 
given by $M/H = \chi/(1+N_c\chi)$, where $N_c\simeq 0.64$ is the demagnetization factor.  Taking $\chi = -0.95$, 
we have $M/H = -2.55$ which is plotted as the broken line in Fig. \ref{squid}b.  
Both the full flux-exclusion implied by $\chi\sim$-1 and the sharp onset are 
evidence for a high degree of electronic homogeneity and uniformity of the condensate.

In low-$T_c$ superconductors, $|M|$ falls steeply above $H_{c1}$, and
continues its monotonic decrease until $H$ attains the upper critical field $H_{c2}$.
Here, the behavior is qualitatively different.  Just above $H_{c1}$, $|M(H)|$ falls to a minimum,
but then slowly increases over several decades in $H$.  Following the initial entry of vortices, the sample 
steadily becomes \emph{more diamagnetic} with increasing $B$.
As we show, this paradoxical pattern reflects the phase-disordering nature of the 
transition at $T_c$ in hole-doped cuprates.

We turn next to temperatures above $T_c$, where the resolution of the
torque cantilever ($\sim 10^{-9}$ emu at 10 T) allows $M$ to be studied in detail.
Figure \ref{MH} displays the $H$ dependence
of $M$ at temperatures increasing from near $T_c$ to 100 K.  Over this broad interval, 
$M$ increases with $H$ with a curvature that becomes increasingly pronounced near $T_c$.  
These curves are strikingly similar to those in the interval $T_s$-$T_c$, 
except for the conspicuous absence of the notch feature associated with a finite $H_{c1}$.

We may combine the SQUID and torque results over our full field and temperature range
in Fig. \ref{MHlog}.  The SQUID results cover the low-$H$ regime (10 Oe to 1000 Oe)
while the torque curves extend to much higher fields.  In this combined plot, $M$
initially increases as a power law in $H$, reaches a broad maximum at fields 1-10 T
and then decreases monotonically.  The fan-like dispersion of the straight lines in low 
fields implies that $M$ has the power-law behavior
\be
M(T,H) = A(T)H^{1/\delta(T)}\quad(H\rightarrow 0),
\label{delta}
\ee
described by a strongly $T$-dependent exponent $\delta(T)$ ($A$ is field-independent).
In the interval 84-105 K, where $\delta > 1$, the non-analytic form of $M$ implies that,
as $H\rightarrow 0$, the diamagnetic susceptibility increasingly exceeds 
linear-response behavior (dashed line drawn for 87 K).


\begin{figure}              
\incl[width=7.5cm]{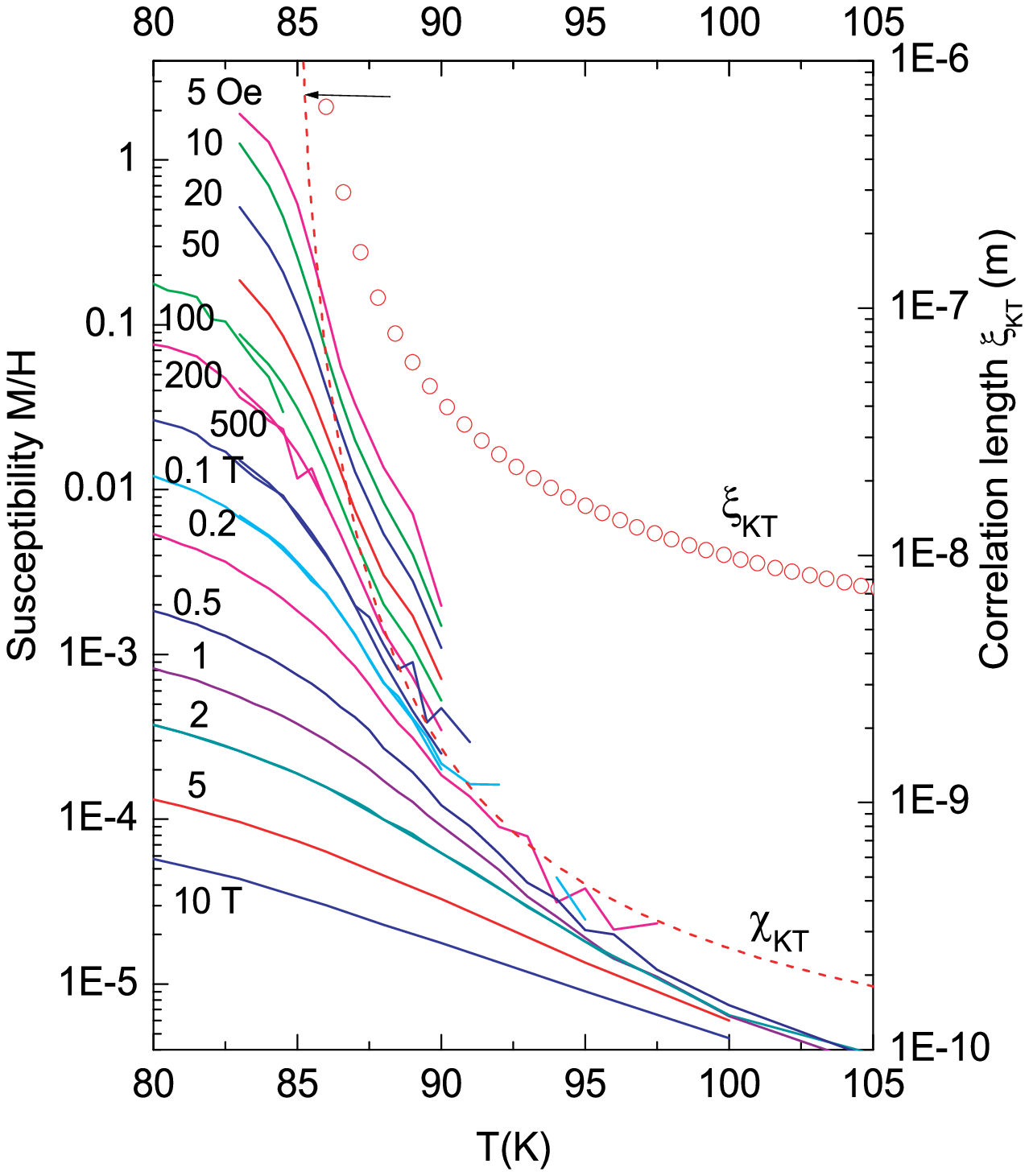}
\caption{\label{chi} The temperature and field dependence of the susceptibility
$\chi(T,H) = M(T,H)/H$.  In weak $H$, $|\chi|$ increases 
by over 5 decades from 100 K to $T_c$, reaching the full flux-expulsion 
value (enhanced by demagnezation) shown by the arrow.  
The dashed curve is a fit to Eq. \ref{MKT} 
with $b = 1$, $d$ = 12 \AA, $a$ = 10 \AA, and $T_{KT}$ = 84 K.  Open circles
represent the KT correlation length $\xi_{KT}$ obtained from the fit.
}
\efig

\bfig	
\incl[width=7.5cm]{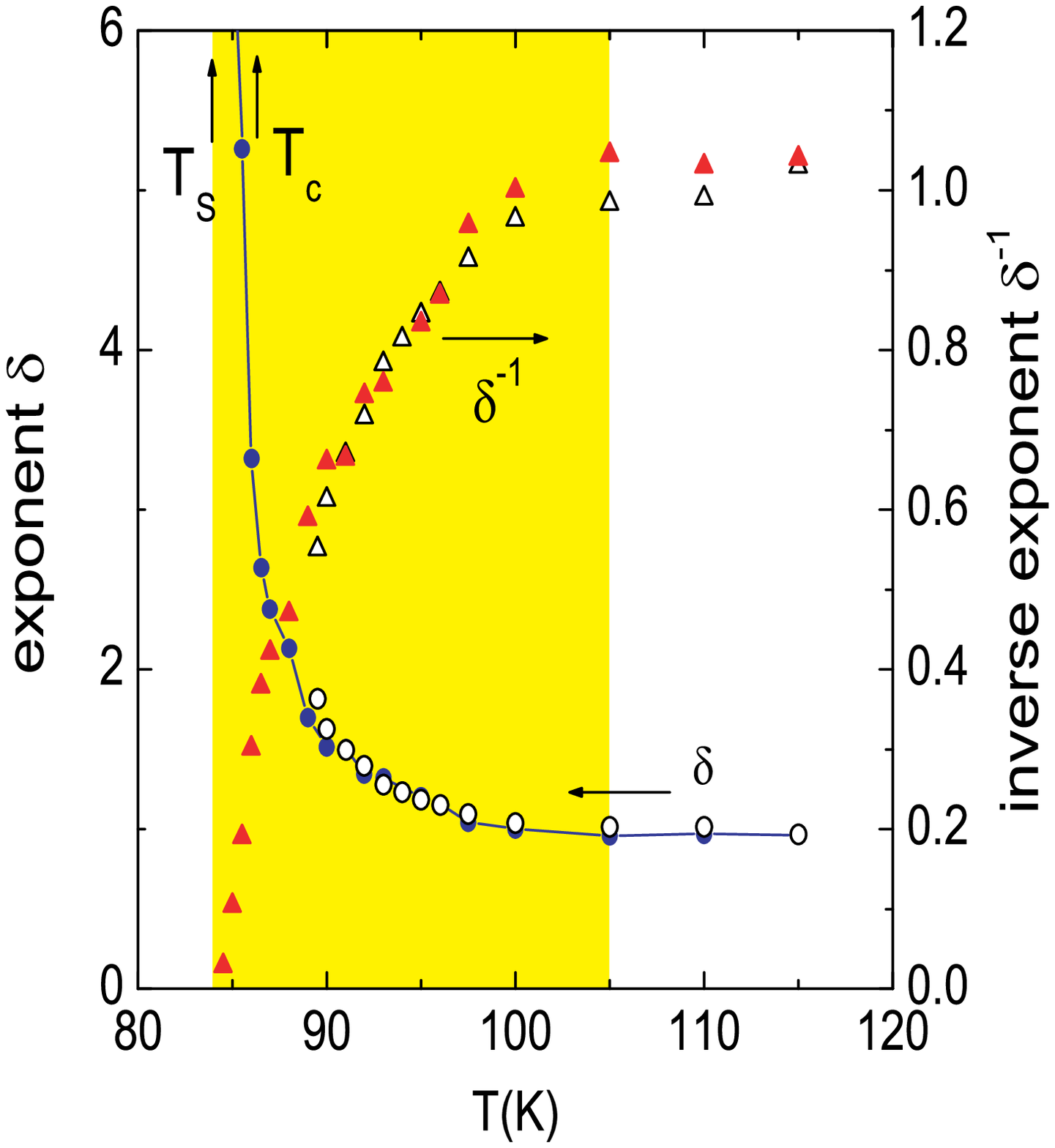}
\caption{\label{expo} The $T$ dependence of the weak-field magnetization exponent $\delta(T)$ 
in Samples I (solid circles) and II (open circles).  The reciprocal 
$\delta(T)^{-1}$ is plotted as solid and open triangles for I and II, respectively.  
As $T\rightarrow T_s^{+}$, $\delta(T)^{-1}$ decreases smoothly to 0.  In the shaded region 
from $T_s$ to 105 K, where $\delta > 1$, linear magnetic response is absent even 
at 10 Oe (Eq. \ref{delta}). 
}
\end{figure}

In Fig. \ref{chi}, the susceptibility $\chi(T,H) \equiv M(T,H)/H$ is displayed versus $T$, with $H$ as the parameter.  
Evidently, $\chi(T,H)$ is highly sensitive to $H$ especially near $T_c$.  
If $\chi$ is measured at a fixed $H$, say 0.5 T, it stays non-divergent across $T_c$.  
However, this is not intrinsic.  Instead, from 105 to 86 K, $|\chi|$ measured in a very weak field 
increases by 5 decades to attain values 0.1--1 just above $T_c$.  As discussed later, this steep increase 
reflects a rapidly changing phase correlation length.

Previous experiments on ``fluctuation diamagnetism'' in cuprates failed to observe a divergent $\chi$
because the $H$ applied was too large~\cite{Vidal,Caretta,Lascialfari}.  The strong $H$ dependence in $\chi(T,H)$ now 
call their conclusions into question.  Further, vortex pinning effects, especially strong in polycrystalline samples
of $\rm{La_{2-x}Sr_xCuO_4}$ and $\rm{YBa_2Cu_3O_{7-\delta}}$, introduce extrinsic features 
in $M$ which render analyses problematical~\cite{Sewer}.

\section{KT correlation length} There are 2 major aspects of the non-linearity in $M(T,H)$ above $T_c$.  The first 
reflects the divergent growth of the phase correlation length $\xi(T)$ while the 
second is the singular behavior in Eq. \ref{delta}.  The 2D phase correlation length measures
the region within which phase coherence prevails in each layer.  In KT theory
long-range phase coherence is destroyed by the appearance of (anti) vortices at the KT transition
(at $T_{KT}$).  Hence the KT correlation length $\xi_{KT}$ is the average spacing between these
thermally created vortices.  Above $T_{KT}$, $M(T,H)$ is linear in $H$ in low fields.  
The weak-field susceptibility is~\cite{Vadim}
\be
\chi_{KT} = -\frac{\mu_0k_BT}{2d\phi_0^2}\xi_{KT}^{2},\quad(T>T_{KT},\; H\rightarrow 0),
\label{MKT}
\ee
where $k_B$ is Boltzmann's constant, $d$ the bilayer spacing, and $\phi_0$ 
the superconducting flux quantum.  The KT correlation length is exponential in the
reduced temperature $t' = T/T_{KT}-1$, viz. $\xi_{KT} = a\mathrm{e}^{b/\sqrt{t'}}$, 
with $b\sim 1$ a non-universal constant and $a$ a cut-off length $\sim$ the vortex core size.  

Clearly, an interlayer coupling $J_{\perp}$ needs to be included to
describe the actual 3D transition at $T_c$ (see below).
Nonetheless, the KT theory describes the essential features of a divergent $\xi(T)$ 
and the roles of thermally-induced vs. field-induced vortices in each layer.  In Fig. \ref{chi}, the dashed curve is a fit 
of Eq. \ref{MKT} with $b$ = 1.0 and $a$ = 10 \AA.  The fit captures quantitatively
the broad features of the 5-decade rise in $\chi(T,H)$ in weak $H$, 
matching its curvature both near $T_c$ and near 100 K.  

The KT correlation length $\xi_{KT}$ inferred from the fit increases 
from $\sim$8 nm at 105 K to 700 nm at 86 K.  A second length 
scale is the spacing $a_B=\sqrt{\phi_0/B}$ of vortices inserted by $\bf H$.
To observe the intrinsic divergence in $\xi_{KT}$, 
we need $a_B\gg\xi_{KT}$ or $B\ll B_{th}$, where $B_{th}\equiv \phi_0/\xi_{KT}^2$ is 
a field-scale set by the density of thermally created (anti)vortices.  
As $T\rightarrow T_s$ in a fixed field $B\ll B_{th}$, $\chi(T)$ initially
increases rapidly (and $B_{th}$ decreases).  However, $\chi$ saturates to a constant
when $B_{th}$ approaches $B$, as evident for the curves in Fig. \ref{chi}.

\section{Singular response as $H\rightarrow 0$} 
Despite the reasonable fit to $\chi(T)$, Eq. \ref{MKT} does not 
account for the non-analytic behavior expressed in Eq. \ref{delta}.  
As mentioned, the deviation between $M(H)$ at 87 K and the dashed line in 
Fig. \ref{MHlog} grows as $H$ decreases. We now 
discuss in more detail the $T$ dependence of this non-analyticity.
From 105 K to 84 K, the weak-field slopes in the log-log 
plots in Fig. \ref{MHlog} decrease smoothly from $\sim$1 to zero, 
i.e. $\delta(T)$ diverges to values exceeding 20.
The steepness of this increase is evident in the plots in 
Fig. \ref{expo} for Samples I and II (solid and open circles, 
respectively).  The \emph{continuous} nature of the increase in $\delta(T)$ 
is made clearer by plotting its reciprocal, which falls below 1 and
smoothly approaches 0 as $T\rightarrow T_s^{+}$ (triangles).  
In Fig. \ref{expo}, the shading highlights the interval from $T_s$ to $\sim$105 K 
in which $\delta$ exceeds 1.  In the shaded interval above $T_c$, $\chi$ diverges without limit
in the limit $H\rightarrow$0 assuming Eq. \ref{delta} remains valid.
Hence, despite the absence of a true Meissner state, a feeble $H$ induces a very large diamagnetic
screening current.  Below $T_c$, however, this limit is interrupted at $H_{c1}$ by the 
appearance of the Meissner state, as discussed in Fig. \ref{squid}b.

Generally, the Meissner state originates from the rigidity of the pair wave function $\hat{\Psi}$.  
In a field, this ``London rigidity" suppresses the paramagnetic current response, leaving only the full 
diamagnetic current~\cite{Schrieffer}.  Here, our magnetization
results imply that, despite the loss of the Meissner effect, $\hat{\Psi}$ 
retains some rigidity above $T_c$.  However, the rigidity is very 
fragile and observable only in a feeble field.  The existence of a singular 
diamagnetic response above $T_c$ has not been predicted (for e.g.,
$\delta$ is strictly 1 above $T_{KT}$ in KT theory).  
Singular behavior in the weak-field vortex-liquid response is discussed in Ref.~\cite{Feigelman}, 
and in Ref.~\cite{Anderson}.  

\section{Discussion}   
We now return to the anomalous behavior of $M(H)$ in the interval $T_s$-$T_c$
where full flux-exclusion prevails for $H<H_{c1}$.  As mentioned, the curves below $T_c$ closely
resemble those above (compare Figs. \ref{squid}a and \ref{MH}).  Aside from fields below $H_{c1}$, 
the curves of $M$ in the log-log plot in Fig. \ref{MHlog} assume a fan-like dispersion that smoothly extends
the behavior in this interval to above $T_c$.  The exponent $1/\delta(T)$, which measures the slopes, 
continuously decreases across $T_c$ (Fig. \ref{expo}).
This continuity strongly implies that the anomalous magnetization in the 2-K 
interval below $T_c$ and above $T_c$ share the same origin; both represent the 
diamagnetic response of a strongly phase-disordered 2D superconducting state with a
very high depairing field $H_{c2}>$ 80 T~\cite{Wang05}.   
Between $T_s$ and $T_c$, the magnetization is comprised of a weak-field 
3D Meissner state that pre-empts the phase-ordering transition of the 2D
condensate at $T_s$.  Application of $H>H_{c1}$ destroys the Meissner state and
uncovers the underlying 2D state whose magnetization continues to 
grow with field.  This accounts for the
unusual ``minimum" in $|M|$ just above $H_{c1}$ mentioned in Fig. \ref{squid}a.  
The steep fall of $|M|$ just above $H_{c1}$ reflects the destruction of the 3D Meissner state, while 
the subsequent power-law increase reflects the behavior of $M$ in 
the underlying 2D condensate which is robust to intense fields.

A hint of our results may be seen in previous torque experiments~\cite{Naughton,Bergemann}.  
In crystals of $\rm{Tl_2Ba_2CuO_{6+\delta}}$ with $T_c$ = 25 and 15 K, the measured curve of $\tau$ vs. $\varphi$ 
displays a puzzling, unexpected term $A_1\sin 2\varphi$ which is actually dominant above $T_c$~\cite{Bergemann}.  
Using Eq. \ref{Meff} and our analysis, we now understand this term to be the robust 2D 
magnetization which persists to intense $H$ and high $T$ as discussed above.  
As noted, the rapid decrease in $|M|$ above $T_c$, compared with the mild change in $\Delta\chi_p$,
engenders a sign change in $A_1(T,H)$ as seen (Fig. 5 in Ref.~\cite{Bergemann}).

The transition in cuprates has been frequently compared to the 3DXY transition with large
anisotropy $\alpha^{-1}$, where $\alpha = J_{\perp}/J$ with $J$ the intralayer coupling 
and $J_{\perp}$ the interlayer coupling~\cite{Janke,Sudbo,Sewer}.  
Our findings that the 2DKT theory gives a broadly accurate description of $\chi(T,H)$ and 
that $T_c$-$T_{KT} \sim \mathrm{2 K} \ll T_c$ confirm that Bi 2212 falls in the 
extreme limit $\alpha\ll 1$.  [By way of comparison, in the planar 
ferromagnet $\rm{K_2CuF_4}$ with $\alpha\ll 1$, the KT theory also describes well
$M$ vs. $H$ above its Curie temperature $T_C$~\cite{Hirakawa}.  However, 
the 3D transition $T_C = 6.25$ K pre-empts the 2D KT transition ($T_{KT}\sim$ 5.6 K)~\cite{Hirakawa,Hikami}.  
Below $T_c$, the destruction of the 3D Meissner state here obviously has no parallel in the planar magnet.]
Unfortunately, the limit $\alpha\ll 1$ is intractable analytically.  
Currently, numerical simulations of $M$ on lattices~\cite{Janke,Sudbo} 
are not accurate enough to compare with our measurements, 
but we hope the detailed measurements here will motivate simulations on larger lattices.

As we cool from 300 K, the first signs of the pair condensate gradually appear at $T_{onset}\sim$ 120 K
as a vortex-Nernst signal~\cite{Xu} and an enhanced diamagnetic susceptibility~\cite{Wang05}.   
The phase of $\hat{\Psi}$ is strongly disordered by vortex motion.
However, the continued growth of these signals to intense 
fields (30-45 T) implies that the depairing field $H_{c2}$ lies much higher (80-100 T).  
Below $\sim$105 K, the non-linear behavior of $\chi$ becomes
increasingly apparent.  In weak $H$, $\chi(T,H)$ undergoes a 5-decade increase as $T_c$ is 
approached, whose broad features can be understood in terms of a diverging phase-correlation length
$\xi$, as prescribed in KT theory (Fig. \ref{chi}).  In the shaded interval in Fig. \ref{expo}, $\chi$ is further enhanced
by a singular power-law dependence suggestive of increasing London rigidity.  

We acknowledge valuable discussions with V. Oganesyan, S. Sondhi, 
V. Muthukumar and P. W. Anderson.  
The research is supported by U.S. National Science Foundation (NSF) through a 
MRSEC grant DMR 0213706 and by the Grant-in-Aid for Science provided by the 
Japan Society for the Promotion of Science. Some measurements were done at the National High Magnetic Field 
Laboratory, Tallahassee, which is supported by NSF and the State of Florida.

\end{document}